\documentclass[doublecol]{epl2}

\title{Observation of the Inverse Cotton-Mouton Effect}
\shorttitle{Observation of the Inverse Cotton-Mouton Effect} 

\author{A. Ben-Amar Baranga\inst{1,\#} \and R. Battesti\inst{1} \and M. Fouch\'e\inst{1,2,3} \and C. Rizzo\inst{1,2,3,*} \and G. L.J.A. Rikken\inst{1}}
\shortauthor{A. Ben-Amar Baranga \etal}

\institute{
  \inst{1} Laboratoire National des Champs Magn\'etiques Intenses\\ (UPR 3228, CNRS-INSA-UJF-UPS), F-31400 Toulouse Cedex, France\\
  \inst{2} Universit\'e de Toulouse, UPS, Laboratoire Collisions Agr\'egats
R\'eactivit\'e, \\ IRSAMC, F-31062 Toulouse, France \\
   \inst{3} CNRS, UMR
5589, F-31062 Toulouse, France\\
\inst{\#} Permanent address : NRCN, P.O.Box 9001, Beer-Sheva 84190, Israel\\
\inst{*} Corresponding author: carlo.rizzo@lncmi.cnrs.fr
}

\abstract{
We report the observation of the Inverse Cotton-Mouton Effect (ICME) i.e. a magnetization induced in a medium by non resonant linearly polarized light propagating in the presence of a transverse magnetic field. We present a detailed study of the ICME in a TGG crystal showing the dependence of the measured effect on the light intensity, the optical polarization, and on the external magnetic field. We
derive a relation between the Cotton-Mouton and Inverse Cotton-Mouton effects that is roughly in agreement with existing experimental data. Our results open the way to applications of the ICME in optical devices.}

\begin{document}

\maketitle

\noindent The Inverse Cotton-Mouton Effect (ICME) is a magnetization induced in any medium by a non resonant linearly polarized light beam propagating in the presence of a transverse magnetic field. This magnetization is proportional to the value of the magnetic field, and to the intensity of the propagating electromagnetic waves (see ref. \cite{Shen} and refs. therein). The ICME was predicted for atomic and molecular systems \cite{Marmo} and for the quantum vacuum \cite{Rizzo}. As stated in ref. \cite{Shen}, microscopically, the light-induced dc magnetization arises because the optical field shifts the different magnetic states of the ground manifold differently, and mixes into these ground states different amount of excited states.

As shown in ref. \cite{Rizzo} the ICME is related to the term in the expansion of the electromagnetic energy of the medium $U$ which is quadratic in the electric and in the magnetic field, which can be written as:
\begin{equation}\label{chiEEBB}
    U = -{1 \over 4} {\epsilon_0 \over \mu_0} \chi_{\alpha \beta \gamma \delta} f_\alpha f_\beta E_\alpha E_\beta B_\gamma B_\delta
\end{equation}
where $\epsilon_0$ is the vacuum permittivity, $\mu_0$ the vacuum permeability, $\chi_{\alpha \beta \gamma \delta}$ is the second order magnetic and electric susceptibility, $f_\alpha, f_\beta$ the local electric field factors, $E$ the electric field and $B$ the magnetic field. Einstein summation is assumed and $_(\alpha, \beta, \gamma, \delta) = x,y, z$. Assuming Kleinman symmetry \cite{Shen},
the medium magnetization can be finally calculated using the relation $M = -{\partial U \over \partial B}$.

A complete experimental proof of the existence of this effect has not been reported yet. In 1987, Zon {\it et al.} reported the measurement of a change in the magnetization of a ferromagnetic film induced by a laser beam in the presence of a static magnetic field parallel to the direction of light propagation \cite{Zon}. This laser-magnetic field geometry is called Faraday configuration and it is usually associated to the Inverse Faraday Effect, not to the ICME as the authors of ref. \cite{Zon} did. The measured magnetization depended on the magnetic field value but not linearly as expected for an ICME. The reported effect did not depend on the laser polarization which is also unexpected.
As far as we know, no other measurement of ICME has been yet reported. For the case of resonant optical
pumping with linearly polarized light in the presence of a magnetic field, measurements of induced magnetization can
be traced back to the sixties \cite{Ziel}. Recently even in the absence of an external magnetic field, a polarization dependent resonant excitation of coherent spin precession by linearly polarized laser pulses has been reported in the antiferromagnet FeBO$_3$ and associated with an optical effect induced by the linearly polarized ultrashort laser pulse acting on spins as an effective field \cite{Kalashnikova}.

The ICME is certainly of interest for applications in optical devices, since it can provide a non demolition method to transform an optical signal into an electric one. Obviously, for applications the effect has to be observed in a medium commonly used in photonics, using a relatively low level magnetic field.
We present here a detailed study of the ICME in a terbium gallium garnet (TGG) Tb$_3$Ga$_5$O$_{12}$ crystal. We chose TGG because is a very common optical crystal, used in particular in Faraday isolators. As expected, our results depend on the laser intensity and polarization, and on the external magnetic field. We also derive a relation between the Cotton-Mouton (CME) and Inverse Cotton-Mouton effects. We compare our prediction with the existing experimental data for the CME of TGG.

The laser source was a Q-switched Nd:YAG laser ($\lambda = 1064$\,nm) providing 10\,ns light pulses of about 0.5\,J/pulse. The laser beam passed through two polarizers. The second one fixed the laser beam polarization, while the first one was used to change the laser power delivered to the TGG crystal. A $\lambda/2$ waveplate was placed behind the polarizers to rotate the laser polarization when needed. Folding mirrors and a lens allowed to deliver and focus the laser beam a few centimeters behind the TGG crystal. Our crystal specifications indicate that damage threshold is $\geq 10^{13}$\,W/m$^2$. As a significant part of our measurements has been done around this value of laser intensity, we performed single shot experiments only, to avoid damage to the crystal. The index of refraction $n$ of TGG at $\lambda = 1064$\,nm is $n \approx 2$, and the crystal absorption at this wavelength is negligible. Crystal dimensions were 2 mm x 2 mm x 2 mm. It was subject to a magnetic field parallel to the [0,1,0] direction provided by an electromagnet. The field values were in the range 0 - 2,5\,T. The ${\bf k}$ vector of light was parallel to the [0,0,1] direction, while the polarization of laser light was parallel to the external magnetic field i.e. parallel to the [0,1,0] direction or perpendicular to the external magnetic field i.e. parallel to the [1,0,0] direction. In the following a subscript $\|$ indicates a quantity measured with the light polarization parallel to the external field, and a subscript $\bot$ indicates a quantity measured with the light polarization perpendicular to the external field.

\begin{figure}[ht]
\label{det}
\includegraphics[width=8cm]{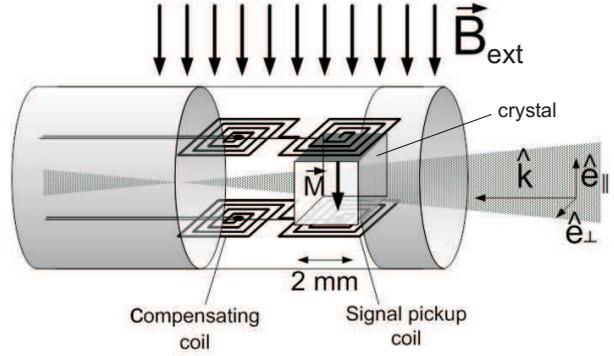}
\caption{Detection apparatus: a laser beam crosses the crystal transversal to an applied magnetic field $B_\mathrm{ext}$. Two sets of pickup-compensating coils monitor the induced crystal magnetization $M$.}\label{Fig:Setup}
\end{figure}

In Fig.\,\ref{Fig:Setup} we show a sketch of the detection zone of the experimental apparatus.

Changes in the crystal magnetization have been measured using a probe constituted by a double pickup coil, a compensating coil and a signal coil.
The signal coil is put in contact with the crystal while the other one is away from the crystal. The double coil is designed in such a way that any signal not coming from the crystal is compensated. Each coil is 2 mm x 2 mm and the distance between the centers of the two coils is 5\,mm. In principle, to avoid signal losses due to returning field lines, the probe should be almost equal in size to the illuminated region and placed as close to it as possible. Each coil has been calibrated by measuring the signal obtained in a known modulated magnetic field. The output signal of the coil is amplified by a low noise fast amplifier and filtered by a 100\,kHz high pass filter. We used two of this type of probes, one for the upper side of the crystal and the other one for the lower side of the crystal. The two probes could be rotated to be sensitive to an induced magnetization parallel or perpendicular to the external magnetic field.

\begin{figure}[ht]
\label{laser}
\includegraphics[width=8cm]{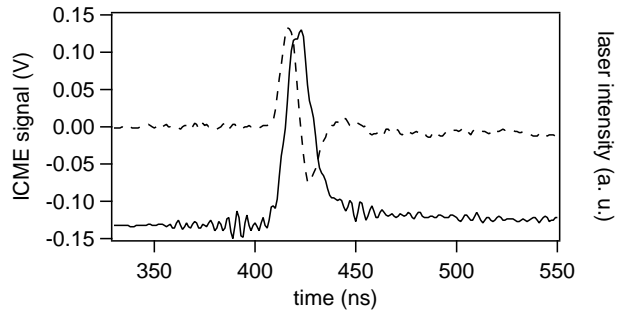}
\caption{Magnetization signal (dashed line) and laser pulse (line) as a function of time. The induced signal follows closely the time derivative of the laser intensity.}\label{Fig:Signals}
\end{figure}

In Fig.\,\ref{Fig:Signals} are plotted a typical laser pulse together with the corresponding signal detected by one of the two signal coils corresponding to an induced magnetization parallel to the external magnetic field. Both signals were recorded on a fast digital oscilloscope with 1\,GS/s. Measurements have been performed for laser polarization parallel and perpendicular to the external magnetic field. At the same value of the external magnetic field, signal intensity depends on light polarization which means that the elements $\chi_{yyyy}$ and $\chi_{xxyy}$ of the $\chi$ tensor defined in Eq. (\ref{chiEEBB}) have two different values. No induced magnetization was observed in the direction perpendicular to the magnetic field. The tensor elements $\chi_{yyyx}$ and $\chi_{xxyx}$ are thus at least negligible compared to $\chi_{yyyy}$ and $\chi_{xxyy}$. For a cubic system like our TGG crystal $\chi_{yyyx}$ and $\chi_{xxyx}$ are expected to be zero for symmetry reasons \cite{Shen}.

The laser pulse is monitored by extracting a small fraction of the beam injected in the crystal with a beam splitter and detecting it with a fast photodiode. The photodiode has been calibrated with respect to an energy meter measuring the pulse energy incident on the crystal.
The ICME signal $V(t)$ is proportional to the time derivative of the magnetic flux through the pickup coil and can be written as:
\begin{equation}\label{Sig}
    V(t)=-g A_\mathrm{e} \frac{dB_\mathrm{p}(t)}{dt},
\end{equation}
where g is the gain of the low noise pickup coil amplifier, $A_\mathrm{e}= 10$\,mm$^2$ is the calibrated effective area of the signal coil and $B_\mathrm{p}$ the average magnetic flux density at the pickup coil position produced by the crystal magnetization $M$. $B_\mathrm{p}$ can be written as:
\begin{equation}\label{Bipi}
    \frac{dB_\mathrm{p}(t)}{dt}=b B_\mathrm{ext} \frac{dI(t)}{dt},
\end{equation}
where $I$ is the laser intensity, $B_\mathrm{ext}$ is the external static transverse magnetic field, and $b$ is a proportionality factor characterizing the ICME. This factor depends on the medium properties and on the pickup coil position with respect to the region of the medium which is illuminated by the laser beam and thus magnetized. Finally, Eq.\,(\ref{Sig}) becomes:
\begin{equation}\label{Sigsig}
     V(t)=-g A_\mathrm{e} b B_\mathrm{ext}\frac{dI(t)}{dt}.
\end{equation}
Therefore, the ICME signal should be proportional to the time derivative of the laser pulse intensity as clearly observed in Fig.\,\ref{Fig:Signals}. We checked that the integrated signal reproduced well the shape of the laser pulse detected by the fast photodiode.

\begin{figure}[ht]
\label{ICME2}
\includegraphics[width=8cm]{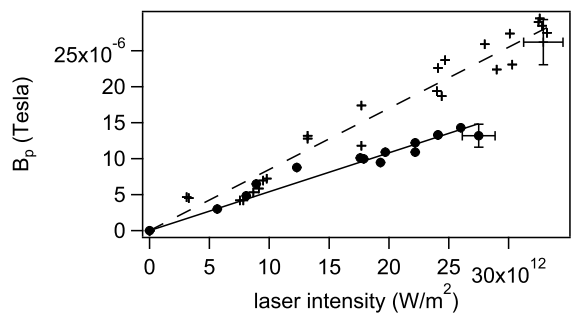}
\caption{Magnetic flux density $B_{\mathrm{p}}$ in the pickup coils versus laser intensity $P_\mathrm{d}$ for two light polarizations: parallel (+) and perpendicular ($\bullet$) to the applied magnetic field. Data are fitted by a linear equation. Error bars represent the typical statistical error.}\label{Fig:Bp_vs_P}
\end{figure}

In Fig.\,\ref{Fig:Bp_vs_P} we show the ICME magnetic flux density at a fixed value of the magnetic field (2.5\,T) varying the laser pulse energy from 0 to 0.25\,J. Data have been taken in two different configurations of the laser polarization: parallel to the external magnetic field corresponding to the measured magnetic flux density $B_{\mathrm{p}\|}$ or perpendicular to the external field corresponding to $B_{\mathrm{p}\bot}$. The diameter of the laser spot in the crystal  was around 1.2\,mm, corresponding to a laser intensity $I$ ranging between 0 and $2.2\times10^{13}$\,W/m$^2$. Fig.\,\ref{Fig:Bp_vs_P} shows that the magnetic flux density depends linearly on the laser intensity as expected. A 12\% statistical error was estimated for the vertical axis and 5\% for the horizontal axis due to pulse to pulse variations in laser energy and uncertainty in laser energy measurement as well as electromagnetic noise induced by Q-switching. For high laser intensity data dispersion can also be ascribed to the proximity to the damage threshold of the crystal.

\begin{figure}[ht]
\label{ICME1}
\includegraphics[width=8cm]{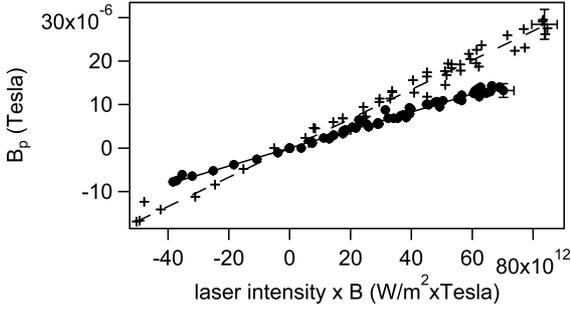}
\caption{Magnetic flux density versus the product of laser intensity by the applied magnetic field for two light polarizations: parallel (+) and perpendicular ($\bullet$) to $B_\mathrm{ext}$. Data are fitted by linear equation. Error bars represent the typical statistical error.}\label{Fig:Bp_vs_PxB}
\end{figure}

In Fig.\,\ref{Fig:Bp_vs_PxB} the complete set of ICME data taken at different values of laser intensity and external magnetic field is plotted. The measured magnetic flux density amplitude is shown as a function of the product of the laser intensity and the external static magnetic field amplitude.

Fig.\,\ref{Fig:Bp_vs_P} and Fig.\,\ref{Fig:Bp_vs_PxB} show that the measured effect depends on light polarization and on the sign of the $B$ field as expected for a real ICME. Together with the correct temporal behaviour of the ICME flux density, this guarantees that we are not hindered by thermo-optic effects like a optical heating driven change in the magnetic susceptibility which could induce a variation of crystal magnetization.

We also show in Fig.\,\ref{Fig:Bp_vs_P} and in Fig.\,\ref{Fig:Bp_vs_PxB} the best linear fit superimposed to the data. The measured magnetic flux density $B_{\mathrm{p}\|}$ and $B_{\mathrm{p}\bot}$ depend linearly on the product of the laser intensity and the external magnetic field value $B_\mathrm{ext}$ as expected, with a proportional factor of $b_\|=(3.36\pm0.04)\times10^{-19}$\,m$^2$W$^{-1}$, and $b_\bot=(2.07\pm0.05)\times10^{-19}$\,m$^2$W$^{-1}$. We have changed the external magnetic field polarity and we have observed that the ICME signal also changed sign. The positive sign of the $b$ constant which means that $B_\mathrm{p}$ is always parallel to $B_\mathrm{ext}$ has thus been verified explicitly.

\begin{figure}[ht]
\begin{center}
\includegraphics[width=5cm]{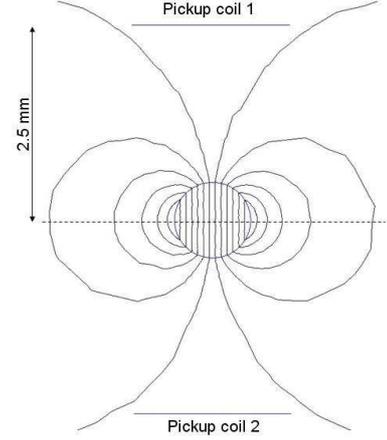}
\caption{Simulation of the magnetic field lines produced by an 1.2\,mm diameter cylindrical magnetization}\label{Fig:Sim}
\end{center}
\end{figure}

To calculate the magnetization of the TGG from the measured value of the magnetic flux density, one has to evaluate the fraction of magnetic flux from the optically magnetized region of the crystal passing through the pick-up coil. Assuming a homogeneous transverse magnetization in a cylindrical region with a diameter of 1.2\,mm, we have calculated the magnetic flux perpendicular to the pick-up coil for our experimental setup, using a finite element magnetic modeler \cite{femm}. In Fig. \ref{Fig:Sim} we show a simulation of the magnetic field lines produced by an 1.2\,mm diameter cylindrical magnetization.

For a magnetization $M$ of 1\,A/m, we calculated a magnetic flux density $B_\mathrm{p}$ of about $4\times10^{-8}$\,T at the position where our signal coil was placed i.e. at about 2.5\,mm from the center of the laser spot. The conversion factor between the measured magnetic flux density $B_\mathrm{p}$ and the crystal magnetization $M$ is therefore about $2.5\times10^7$\,(A/m)T$^{-1}$.

The ICME magnetization of the TGG can be defined as:
\begin{equation}\label{Magn}
    M = C_\mathrm{ICM} I B_\mathrm{ext}
\end{equation}
where $C_\mathrm{ICM}$ depends only on the medium properties and therefore one may call it the Inverse Cotton-Mouton constant. Our data indicates therefore that for our TGG crystal $C_{\mathrm{ICM}\|}=8.4\times10^{-12}$\,(A.m)(W.T)$^{-1}$ and $C_{\mathrm{ICM}\bot}=5.2\times10^{-12}$\,(A.m)(W.T)$^{-1}$.

As far as we know, no theoretical prediction of the value of the ICME in a TGG crystal exists.
From Eq. (\ref{chiEEBB}), since $f_\alpha = f_\beta = f$ in this case, $\Delta M \equiv M_\| - M_\bot$ can  be expressed as:
\begin{equation}\label{DeltaM}
    \Delta M = {1 \over 2}{\epsilon_0 \over \mu_0} \left(\chi_\| - \chi_\bot\right) f^2 E^2 B_\mathrm{ext} \equiv {\Delta C_{\mathrm{ICM}}} I B_\mathrm{ext},
\end{equation}
where $\chi_\| = \chi_{yyyy}$ and $\chi_\bot = \chi_{xxyy}$ and $\Delta C_{\mathrm{ICM}} \equiv C_{\mathrm{ICM}\|} - C_{\mathrm{ICM}\bot}$.

On the other hand, the Cotton-Mouton effect is related to the variation of the index of refraction induced by the transverse magnetic field $\Delta n_\mathrm{CM} \equiv n_\| - n_\bot$ \cite{RizzoRizzo}. Following the quantum theory developed for media with $n\neq 1$ \cite{Cappelli}, one can write:
\begin{equation}\label{DeltaEnne}
    \Delta n_\mathrm{CM} = {1 \over 4 \mu_0 n} \left(\chi_\| - \chi_\bot\right) f^2 B_\mathrm{ext}^2\equiv k_\mathrm{CM}B_\mathrm{ext}^2,
\end{equation}
where $n$ is the index of refraction without magnetic field.

Eqs.\,(\ref{DeltaM}) and (\ref{DeltaEnne}) show that a simple relation exists between $k_\mathrm{CM}$ and $\Delta C_{\mathrm{ICM}}$ :
\begin{equation}\label{ratio}
\frac{k_\mathrm{CM}}{\Delta C_{\mathrm{ICM}}} = {c \over 2n},
\end{equation}
where $c$ is the velocity of light.
We can therefore estimate the value of the Cotton-Mouton effect for our TGG crystal from our ICME measurement: $k_\mathrm{CM} \approx 10^{-4}\,$T$^{-2}$. As far as we know the only experimental value for the CME of a TGG crystal is reported in ref. \cite{Kolmakova}. The values reported in Fig. 1 of ref. \cite{Kolmakova} are of the order of 10$^{-4}$, for a 4 T magnetic field, at low temperature ($\leq$ 50 K), for a different laser wavelength ($\lambda = 0.63$\,$\mu$m) and different configuration of the ${\bf k}$ vector of light with respect to crystal orientation. The comparison between our prediction based on Eq. (\ref{ratio}) and the values reported in \cite{Kolmakova} is not straightforward since the experimental parameters are very different, nevertheless it looks that reported values of the CME are roughly in agreement with our predicted value. New measurements of the CME of TGG are necessary to accurately test the validity of our Eq. (\ref{ratio}).

Our observation of the ICME in a crystal of TGG opens a new field of investigations of electromagnetic properties of matter since the CME gives only information on the difference between matter response in particular elements of the $\chi$ tensor defined by Eq. (\ref{chiEEBB}) while the Inverse Cotton-Mouton effect provides an absolute value for this response, in our case the tensor elements $\chi_{yyyx}$, $\chi_{xxyx}$, $\chi_{yyyy}$ and $\chi_{xxyy}$.

Our result is also encouraging to push forward ICME studies in dilute matter as discussed in ref. \cite{Rizzo}.
We also show that ICME provides a non demolition method to transform an optical signal to an electric one, which is of interest for application in optical devices \cite{pat}. Further studies are certainly necessary to optimize the present TGG crystal configuration or to look for other materials in view of applications.

We thank Anna Mamaliga and the technical staff of the LNCMI, in particular G\'eraldine Ballon. This work is supported by \textit{EuroMagNET} and \textit{Fondation pour la recherche IXCORE}.

\end{document}